\documentclass[prb,twocolumn,aps,showpacs]{revtex4}
\usepackage{graphicx}
\usepackage{amsfonts}
\usepackage{bm}

\newcommand{\app}{\;\!_\circ^\bullet\;\!_\circ^\bullet}
\newcommand{\bp}{\;\!_\bullet^\bullet\;\!_\circ^\circ}
\newcommand{\cp}{\;\!_\circ^\bullet\;\!_\bullet^\circ}
\newcommand{\dpp}{\;\!_\bullet^\circ\;\!_\circ^\bullet}
\newcommand{\ep}{\;\!_\circ^\circ\;\!_\bullet^\bullet}
\newcommand{\fp}{\;\!_\bullet^\circ\;\!_\bullet^\circ}

\begin{document}
\title{Effects of many-electron jumps in relaxation and conductivity 
of Coulomb glasses}
\author{J. Bergli}
\affiliation{Department of Physics, University of Oslo, P.O.Box 1048
  Blindern, N-0316 Oslo, Norway}
\author{A.\ M.\ Somoza and M.\ Ortu\~no}
\affiliation{Departamento de F\'{\i}sica - CIOyN, Universidad de Murcia, 
Murcia 30.071, Spain}

\begin{abstract}
A numerical study of the energy relaxation and conductivity of the
Coulomb glass is presented. The role of many-electron transitions is
studied by two complementary methods: a kinetic Monte Carlo algorithm
and a master equation in configuration space method. A calculation of
the transition rate for two-electron transitions is presented, and the
proper extension of this to multi-electron transitions is discussed.
It is shown that two-electron transitions are important in bypassing
energy barriers which effectively block sequential one-electron
transitions. The effect of two-electron transitions is also discussed.
\end{abstract}
\maketitle

\section{Introduction}

At low temperatures, disordered systems with localized electrons
(e. g. on dopants of compensated doped semiconductors or Anderson
localized states in disordered samples) conduct by phonon assisted
hopping. The theory of this process goes back to Mott \cite{mott} who
invented the concept of variable range hopping. In particular he
derived the Mott law for the temperature dependence of the conductance
\[
\sigma \sim e^{-(T_0/T)^{1/(d+1)}}
\]
where $d$ is the dimensionality of the sample. If Coulomb interactions
are important one describes the system as a Coulomb glass due to the
slow dynamics at low temperatures. As is well known (See
Ref. \onlinecite{ES} and references therein), the single particle
density of states develops a soft gap at the Fermi level, the so
called Coulomb gap. While this understanding of the density of states
is generally accepted, the situation is less clear when it comes to
describing dynamics in the interacting case. Using the Coulomb gap
density of states, which is a result of interactions, in the variable
range hopping argument, assuming that it can be used in the same way
as the non-interacting density of states, yields the Efros-Shklovskii
law for conductance
\[
\sigma \sim e^{-(T_0/T)^{1/2}}.
\]
While this has
been observed experimentally in several cases it is not always the
case. It remains clear that this is an uncontrolled approximation, and
the full theoretical understanding of this is still missing.

In particular, since this approach is based on a single-particle
approximation, it neglects the possibility of correlated jumps of two
or more electrons. At low temperatures, the system can be trapped in
metastable configurations, from which it can be difficult to escape by
single-electron transitions. By a two-electron transition the system
can jump out of this metastable state even when the temperature is so
low that the probability of making the same transition sequentially is
very small because it passes through an activated intermediate state
with higher energy. Thus, one would expect the importance of
many-electron jumps to increase as the temperature is decreased. This was also
the conclusion of some works \cite{tenelsen, pg} which used a method
which identifies the full set of low energy states of the system, and
studies the possible transitions between them. Because the number of
accessible states grows rapidly with increasing temperature and system
size, this method is restricted to small systems and low
temperatures. 

The importance of many-electron jumps was disputed by
Tsigankov and Efros \cite{tsigankov}, who used a kinetic Monte Carlo
method to study the dynamics of the Coulomb glass. Using only
one-electron transitions, they confirm the Efros-Shklovskii law both
regarding the value $\frac{1}{2}$ in the exponent and also the value
of $T_0$ predicted by percolation theory\cite{ES}. Including
two-electron transitions, they find that the two-electron jumps
contribute about two orders of magnitude less to the current than the
one-electron jumps. Furthermore, they find that the relative
contribution of two-electron jumps decreases with decreasing
temperature. They conclude that two-electron jumps are not important
for the conductance of the Coulomb glass, contradicting the previous
works \cite{tenelsen, pg}. We do not understand how their results can
lead to this conclusion, since the two-electron jumps can be crucial
in facilitating transport through one-electron jumps, even if their
actual number is much less than the number of one-electron jumps. What
should be compared is the conductance when only allowing one-electron
jumps with the conductance when two-electron jumps are also
included. This will be discussed in more detail later.  

Tsigankov and Efros \cite{tsigankov} explain their contradiction with
the previous works \cite{tenelsen, pg} as coming from two sources:
First, in Ref. \onlinecite{pg} the rates of two-electron transitions
were overestimated because they were assumed to be independent of the
distance between the two electrons involved in the transition. This
assumption is not reasonable, since it is the Coulomb interaction
which allows the remote electrons to exchange energy, and the
probability of a double jump should decrease with distance. Second,
the method of identifying the full set of low energy states used in
Refs. \onlinecite{tenelsen, pg} is numerically costly, and therefore
limited to very small samples and low temperatures (where only the
states with the very lowest energies are thermally excited). Therefore
the conclusions of Refs. \onlinecite{tenelsen, pg} may be the result
of small sizes and not valid for larger systems. An attempt to answer
the second criticism was made in Ref. \onlinecite{somoza} where the
size dependence was studied and a scheme for extrapolation to infinite
size was suggested. Using the same percolation method in configuration
space as in Ref. \onlinecite{pg} (including the same expression for
the many-electron transition rate which was questioned in
Ref. \onlinecite{tsigankov}) it was still found that many-electron
jumps become important at low temperatures. The difference with the
results of Ref. \onlinecite{tsigankov} is explained by the fact that
the method they used, studying all transitions between an extensive
set of low energy states and involving up to six electrons moving
together, was more suited to identify the crucial many-electron
transitions.  These could enhance the conductance even if they happen
only very rarely since they could facilitate subsequent one-electron
transitions. Since the Monte Carlo method\cite{tsigankov} becomes
impracticably slow at low temperatures while the percolation in
configuration space\cite{pg,somoza} only can be applied for small
systems and low temperatures, several questions remained unanswered:
Is the difference in the expression for the many-electron transition
rate the reason for the difference in results? Are the two methods
equivalent, or does one of them contain systematic errors? At what
temperatures are the many-electron transitions important?

In this work we try to answer these questions. Previous works have
focused on the influence of multi-electron transitions on conductance
since this is the commonly measured quantity. However, at low
temperatures it can be difficult to numerically find the conductance
for two reasons: First, the system should be equilibrated which is a
slow process at low temperatures. Second, to be in the linear response
regime one must use a small potential difference across the sample,
and the resulting current is so weak that one needs a long sampling
time to get accurate values. Therefore we have chosen to study the
relaxation of energy instead, this being a quantity easily accessible
in simulations. It is also well known that
experimentally\cite{zvi,grenet} the conductance is also slowly
relaxing, so that relaxation may be as important and relevant as
equilibrium conductance (Although one should keep in mind that the
experiments of Ref. \onlinecite{grenet} are on granular aluminium
films, and it may be that this system is more complex than the model
discussed here accounts for). This allows us to go to lower
temperatures using the Monte Carlo method, and thereby bridge the gap
to the configuration space method. Here we report on the following: We
give a direct calculation of the transition rate for two-electron
jumps, to replace the unphysical one used in Refs. \onlinecite{pg,
somoza} and the approximate one suggested in
Ref. \onlinecite{tsigankov}. A preliminary report of this part of our
work was already presented in Ref. \onlinecite{BeSo09}. We also
discuss the extension of this result to many-electron transitions
(Sec. \ref{transitionRates}). We have numerically studied the
relaxation of the total energy of the system, comparing the evolution
when only allowing one-electron jumps with the one where two- and
three-electron jumps are included. We have used both the Monte Carlo
algorithm suggested by Tsigankov and Efros\cite{tsigankov} and the
configuration space method of Refs. \onlinecite{pg, somoza}, but
instead of using the percolation method we have studied the master
equation on the set of low energy states, thereby eliminating any
doubt on the accuracy of the percolation method. The details of the
models used and the numerical procedures are given in
Sec. \ref{model}. The results on energy relaxation are presented in
Sec. \ref{results} and some results on conductance in
Sec. \ref{conductivity}.

\section{Many-electron transition rates}\label{transitionRates}

We start from the standard Coulomb gap Hamiltonian with a perturbation
term due to tunneling
\begin{equation}
 H_0 = \sum_i\phi_i c_i^\dag c_i+\sum_{i<j} V_{ij}c_i^\dag c_ic_j^\dag c_j
 +\sum_{i<j}t_{ij}c_i^\dag c_j + \mbox{h.c.}
\end{equation}
describing localized electrons interacting through Coulomb forces.
$c_i^\dag$ and $c_i$ are operators creating and annihilating an
electron on site $i$, $\phi_i$ is the intrinsic energy of site $i$,
which we assume to be a random variable uniformly distributed in the
interval $[-W/2,W/2]$, and $V_{ij}=e^2/{r_{ij}}$ is the Coulomb
energy.  The tunneling amplitude $t_{ij}=I_0 e^{-2r_{ij}/a}$ depends
exponentially on the distance $r_{ij}$, and $a$ is the localization
radius and the prefactor is $I_0=e^2/\kappa a$ with $\kappa$ the
dielectric constant.

We consider phonon assisted tunneling 
due to the electron-phonon interaction:
\begin{equation}
  H_{\text{e-ph}} = \sum_{\bf q} \sum_i c_i^\dag c_i \left( e^{-i{\bf
      qr}_i}\gamma_{\bf q}b_{\bf q} + \mbox{h.c.} \right)
\end{equation}
where $b_{\bf q}$ is the phonon annihilation operator and $\gamma_{\bf
  q}$ is a numerical factor depending on the exact phonon
interaction. 

The one-electron transition rate from site $i$ to site $j$ is
well known, see for example Ref. \onlinecite{ES},
\[
\Gamma_{ij} \propto |\gamma_q|^2N(\Delta E)e^{-2 r_{ij}/a}
\]
where $r_{ij}$ is the distance between the sites and $\Delta E$ is the
change in energy.  $N(E) = 1/(e^{E/T}-1)$ is the equilibrium phonon
density, for emmision processes this has to be replaced by
$N(E)+1$. We set $k_B=1$ so that temperatures and energies are
measured in the same units. In this work, following
Ref. \onlinecite{tenelsen} we use the  formula
\[
\Gamma_{ij} = {\tau_0}^{-1}e^{-2 r_{ij}/a}\min 
  \left(e^{-\Delta E/T},1\right)
\]
where $\tau_0$ contains material dependent factors and energy
dependent factors, which we approximate by their average value; we
consider it as constant and its value, of the order of $10^{-12}$ s is
chosen as our unit of time (Note that in Ref. \onlinecite{tsigankov} a
different formula was used, we do not believe that the difference is
of great significance, although it may change numerical values).

\subsection{Two electron transition rates}

Let us concentrate for the moment in transitions of two electrons on
four sites and follow the method described in Ref.
\onlinecite{BeSo09}, based on the locator expansion in configuration
space.  We can restrict ourselves to the four sites involved in the
transition and include the Coulomb interaction with the rest of the
system in the site energy $\phi_i$. The zero-order (in the tunneling
perturbation) configurations of two electrons on four sites are
described by the states
\begin{equation}
 |\app\rangle\quad |\bp\rangle \quad
 |\cp\rangle\quad
 |\dpp\rangle \quad  |\ep\rangle  \quad
 |\fp\rangle 
\end{equation}
where filled circles represent sites with electrons and empty circles
empty sites.
 The sites are numbered as $|\;\!_3^1\;\!_4^2\rangle$.
We calculate the initial and the final states to second order in the tunneling,
and we denote them by $|\widetilde \app\rangle$ to $|\widetilde \fp\rangle$. 

We consider phonon assisted tunneling from the initial perturbed
state $|\widetilde \app\rangle$ to the final state $|\widetilde \fp\rangle$. 
The calculation is a direct generalization of the one found in
Ref. \onlinecite{ES} for one-electron transitions. We assume that the
electron on an impurity is described by a hydrogen-like wavefunction
with a localization radius $a$ and that $qa\ll1$ where $q$ is the
wave vector of the phonon, so that 
\begin{equation}
\langle i|e^{-i{\bf qr}_i}|i\rangle  \approx e^{-i{\bf qr}_{i}}
\end{equation}
where $|i\rangle$ describes an electron on impurity $i$.
After some algebra we find
\begin{eqnarray}
 M_{\bf q} &
=\langle\widetilde \app|\sum_ie^{-i{\bf qr}_i}c_i^+c_i|\widetilde \fp\rangle
 = t_{13}t_{24}\frac{E_{\fp}-E_{\dpp}
     +E_{\app}-E_{\cp}}{E_{\fp}-E_{\app}}\nonumber\\
  &\left[
    \frac{e^{-i{\bf qr}_1}
      -e^{-i{\bf qr}_3}}{(E_{\app}-E_{\cp})(E_{\fp}-E_{\dpp})}
   +\frac{e^{-i{\bf qr}_2}
      -e^{-i{\bf qr}_4}}{(E_{\app}-E_{\dpp})(E_{\fp}-E_{\cp})}
  \right]
\label{matrix}\end{eqnarray}
$E_\alpha$ refers to the energy of configuration $|\alpha\rangle$.
This expression corresponds to electrons jumping from site 1 to 3 and
2 to 4.  We have to add a similar expression involving the jump from 1
to 4 and from 2 to 3. If the sites are at random positions, the jump
with the minimum total hopping distance will dominate and we can
neglect the other jumping possibilities, but if the sites are on a
lattice we have to keep all the terms (and their cross terms) in the
calculation of $|M_{\bf q}|^2$.

We further assume that $qr_{ij}\gg 1$ (this may fail at sufficiently
low temperatures), which allows us to replace factors of the form
$\cos{\bf q r}_{ij}$, appearing in matrix elements of wavefunctions of
different sites, by 0 when integrating over the directions of ${\bf
q}$.  Let us concentrate on the different energy factors appearing in
Eq.\ (\ref{matrix}).  We first note that $E_{\fp} -E_{\app} =\Delta E$
is the total energy difference and will be equal to the energy of the
phonon that is emitted or absorbed which then determines the phonon
wave vector. Further, 
\begin{equation}
E_{\fp}-E_{\dpp}+E_{\app}-E_{\cp}\sim
V_{12}+V_{34}-V_{14}-V_{23}\equiv V_{13,24}
\label{dipoledipole}\end{equation}
is independent of the site energies. It only depends on the
geometrical disposition of the jumps and if the separation between
sites of different jumps is much larger than both jumping distances it
corresponds to the dipole-dipole interaction.  The energy denominators
in Eq. (\ref{matrix}) involve the intermediate states, and it is very
CPU time consuming to calculate these terms in numerical simulations.
The divergence of these terms at certain points reflect the limitation
of the perturbation theory rather than any physical effect. Therefore
we want to cut off this divergence and replace the fraction by 1 when
it is larger than 1. The energy differences are of the order of the
disorder $W$.  Therefore the terms in the brackets are also never very
small. Since they are also temperature independent we propose to set
these terms equal to $1/W$. We believe that this is of no physical
consequence, and will not affect the results qualitatively.  Taking
into account these approximations and using Fermi's golden rule we
arrive at the expression for the transition rate
\begin{equation}\label{two}
\Gamma_{\app\rightarrow\fp}
=\tau_0^{-1}\frac{I_0^2 V_{13,24}^2}{(W/2)^4}e^{-2(r_{13}+r_{24})/a}
\min\left(e^{-\Delta E/T},1\right)
\end{equation}
where $\tau_0$ is the same unit of time as for one-electron
transitions and we assume that the average energy difference of
intermediate states is W/2. More details on the derivation of this
equation were given in Ref. \onlinecite{BeSo09}.

\subsection{Three and more electron transition rates}

In this case one has to assume that the interaction is weak and do
perturbation theory in both the interaction and the hopping term. As
for two-electron jumps, it is an excellent approximation for sites at
random to consider only the transitions with the minimum total hopping
distance $R_{I,J}$. There are many of them, differing in the order of
the one electron moves and on the jump directly excited by the phonon.
The final expression for the transition rate is very complex, but its
most important factor is easy to get \citep{Po81}.  It is the 
probability of finding a phonon of energy $\Delta E$ times the product
of the overlap integrals of the hops of all electrons involved in the
transition.  The many electron transition rate can then be
approximated by
\begin{equation}\label{manytr}
\Gamma_{I,J}
= \tau_0^{-1}\gamma^{n-1} e^{-{2R_{I,J}}/{a}}\min 
  \left(e^{-\Delta E/T},1\right).
\end{equation}
$\gamma$ is a measure of the importance of the interaction energy
compared to the disorder energy and $n$ is the number of electrons
participating in the process.

The use of equation (\ref{manytr}) for the transition rates in
numerical simulations may overestimate the importance of correlated
hops since it doublecounts the effects of excitations well separated
one from each other. Although one-electron excitations should dominate
in this case, since many-electron excitations should only be important
when single excitations have positive energies, while the combined
excitation has negative energy, it is convenient to get rid of this
problem.  We can do that by substituting the constant $\gamma$ by a
prefactor similar to the one obtained for two-electron transitions,
equation (\ref{two}). It is difficult to get a closed expression
for this prefactor, and we propose an empirical approach that it is
practical for numerical purposes and that we think incorporates the
relevant physics of the problem. A requirement for this prefactor is
that it should vanish when one of the transitions is very far from the
others. A suggestion satisfying this requirement is the sum of all the
products of $n-1$ different interaction energies between any pair of
single electron transitions, like $V_{13,24}$ in
(\ref{dipoledipole}). Each of these terms must be divided by a factor
proportional to the disorder energy as in the two-electron case. This
proposal corresponds to exciting one of the hops by the phonon and the
rest by the dipole-dipole interaction in all the possible ways.
For three electron transitions we take as the
preexponential
\begin{equation}
\frac{I_0^2}{\tilde{W}^6}(V_{14,25}^2 V_{25,36}^2 +V_{25,36}^2
V_{14,36}^2 +V_{14,25}^2 V_{14,36}^2).
\end{equation}

\section{Model for numerical simulations}\label{model}

We use the standard tight--binding Coulomb gap Hamiltonian \cite{ES}:
\begin{equation}\label{hamil}
H=\sum_i\epsilon_i n_i +\sum_{i<j} {\frac{(n_i-K)(n_j-K)}{r_{ij}}}\;, 
\end{equation}
$K$ being the compensation.  We take the number of electrons to be
half the number of sites.  The sites are arranged in two dimensions
both on a lattice and at random, but in the latter case with a minimum
separation between them, which we choose to be $0.05r$ where
$r^2=L^2/N$.  We implement cyclic boundary conditions in both
directions.  We take $e^2/r$ as our unit of energy and $r$ as our unit
of distance.

\subsection{Monte Carlo algorithm for lattice systems}

For the two-electron transition rate we use Eq.\ (\ref{two}) for sites
at random and the extension that includes the two jumping
possibilities when sites are on a lattice.  As for one-electron
transitions, the rate is split in one energy dependent (or activation)
term, $\Gamma^A$ and one distance dependent (or tunneling) term
$\Gamma^T$. This means that we can use the hybrid algorithm of
Tsigankov and Efros \cite{tsigankov}.

The program first calculates and stores the distance dependent part of
the rates. For the one electron jumps, the tunneling parts of the rate
for all jumps in the square
$|\Delta x|,\; |\Delta y| \leq L_1$ some maximal jump
length (in the numerics $L_1=10$ lattice units) are
calculated. The sum $\Gamma_{T,1}^{Total} = \sum \Gamma_{T,1}$
is also stored. For the two electron jumps, all coordinates are relative
to the initial position of the first electron. The following algorithm
is used: 

\begin{itemize}
\item{The final position of the first electron is selected in the
    square $|\Delta x|,\; |\Delta y| \leq L_2$ where the size $L_2$
    can be reasonably chosen to be about half of $L_1$ since the
    distances each electron jumps are added together to find the
    rate (In the numerics $L_2=3$ lattice units). 
}
\item{The initial position of the second electron is selected in the
    square $|x_2|,\; |y_2| \leq D_2$ (In the
      numerics $D_2=5$ lattice units). The initial position of the second
    electron can not be either the initial or the final position of
    the first electron.}
\item{The final position of the second electron is selected in the
    square $|\Delta x_2|,\; |\Delta y_2| \leq L_2$. The final position
  of the second electron can not be the initial or final position of
  the first electron.}
\item{The tunneling part of the rate for this transition is calculated
    according to the formula 
    \[
    \Gamma_{T,2} = E_1^2V_{13,24}^2 + E_2^2V_{14,23}^2 
         + E_1E_2V_{13,24} V_{14,23}
    \]
    where
    \[
    E_1 = e^{-2(r_{13}+r_{24})/a}, \qquad E_2 = e^{-2(r_{14}+r_{23})/a}
    \]
    }
\item{The rates for all these transitions are stored, and the sum of
    all $\Gamma_{T,2}^{Total} = \sum \Gamma_{T,2}$ is calculated.}
\end{itemize}

When the Monte Carlo algorithm is running, it will do the following
steps in order to select which transition to make. 
\begin{itemize}
\item{It is decided whether to attempt a one electron transition
    (probability
    $\Gamma_{T,1}^{Total}/(\Gamma_{T,1}^{Total}+\Gamma_{T,2}^{Total})$)
  or a two electron transition
    (probability
    $\Gamma_{T,2}^{Total}/(\Gamma_{T,1}^{Total}+\Gamma_{T,2}^{Total})$)
}
\item{If it is a one electron transition we follow the usual procedure
    of Tsigankov and Efros\cite{tsigankov}. One occupied site is selected
    randomly. Then the final site is selected at random but with
    weights given by the probability
    $\Gamma_{T,1}/\Gamma_{T,1}^{Total}$. If the final site is
    occupied, the transition is rejected. If it is empty it is
    accepted with probability $\min(1,e^{-\Delta E/T})$.
}
\item{If it is a two electron transition an occupied site is selected
    at random. Then a certain two electron jump is selected weighted
    by the  probability $\Gamma_{T,2}/\Gamma_{T,2}^{Total}$. If the
    final site of the first electron is occupied, the initial site of
    the second electron is empty or the final site of the second
    electron is occupied, the transition is rejected. If not, the
    transition is
    accepted with probability $\min(1,e^{-\Delta E/T})$.
}
\end{itemize}

\subsection{Master equation method for sites at random}

We used a numerical algorithm to obtain the ground state and $\approx
10^5$ lowest energy many--particle configurations of the system.  The
algorithm is an improved version of the algorithm in Ref.\
\onlinecite{somoza} and it was described in detail in
\onlinecite{ds}.  It consists of the following two stages.  In the
first stage, we repeatedly start from states chosen at random and
relax each sample by means of a local search procedure.  In an
iterative process, we look for neighbors of lower energies and always
accept the first such state found. The procedure stops when no lower
energy neighboring states exist, which insures stability with respect
to all one--electron jumps and compact two--electron jumps.  We then
consider a set of metastable sates found by the process just described
and look for the sites which present the same occupation in all of
them. These sites are assumed to be frozen, {\it i.e.} they are not
allowed to change occupation, and the relaxation algorithm is now
applied to the unfrozen sites. The whole procedure is repeated until
no new frozen sites are found with the set of metastable states
considered.

In the second stage, we complete the set of low--energy configurations
by generating all the states that differ by one-- or two--electron
transitions from any configuration stored.
In order to speed up this process, which is very CPU time consuming,
we again assume frozen and unfrozen sites and in first place look
for neighboring configurations by changing the occupation of unfrozen
sites only. We later relax this restriction in the final stage.

We consider 20 different realizations of a system with 500  sites.
We obtain the 200 000 lowest energy configurations for each realization.
We then obtain all one- two- and three electron transitions between all 
these configurations, establishing a dynamical matrix that evolves the 
system in time according to this master equation. 
We choose for the initial state the mixture of all configurations
with weights equal to the Boltzmann factor for a temperature 40 times 
larger than the real one.

We have developed a renormalization procedure to propagate in time
efficiently.  It takes advantage of the fact that the distribution of
relaxation times is exponentially broad and at large times the short
time processes must have already equilibrated some sets of
configurations.  At a given time, we calculate the current through
each transition and if it is very small, relative to the transition
rate and the occupation probability, we assume that the two
configurations involved are in thermal equilibrium and can consider
then as part of a cluster. We recalculate the transition rates between
this new cluster and the rest of the system.  The time step used in
the numerical propagation is calculated dynamically and increases
drastically as we form more clusters or larger clusters.

\section{Results on energy relaxation}\label{results}

\subsection{Relaxation using Monte Carlo on the lattice model}

To see the effect of two electron transitions on relaxation we did the
following. For one sample of size $100\times100$ and $W=2$ we ran
relaxation from an initial random state at three different
temperatures, $T$ = 0.01, 0.005 and 0.001. For each temperature we ran
from the same initial state 10 (20 for $T=0.001$) different time
evolutions (different random seeds for the Monte Carlo evolution). For
each case we ran the simulation both allowing only one electron
transitions and including one and two electron transitions.
\begin{figure*}  
\includegraphics[width=.3\linewidth]{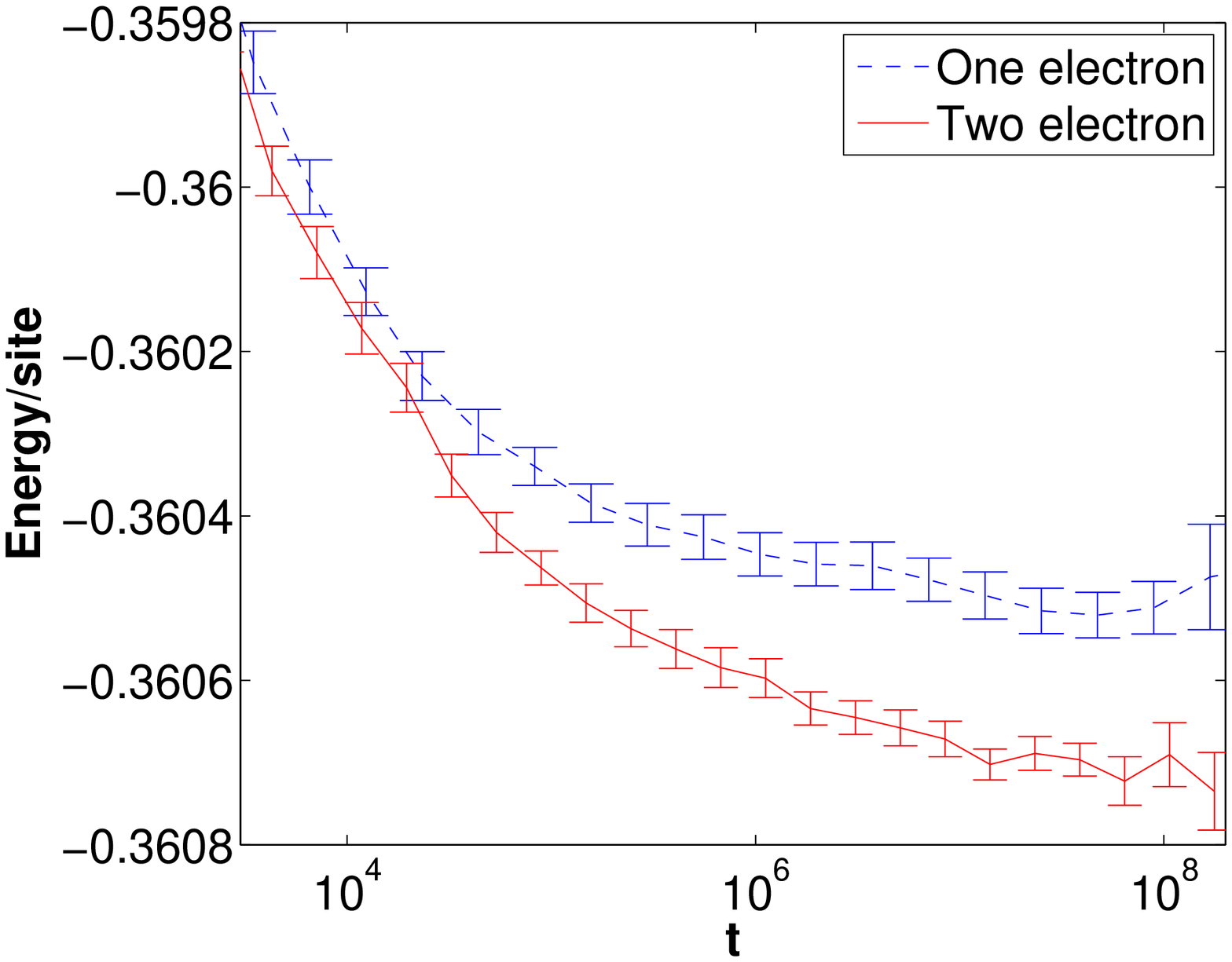}a)
\includegraphics[width=.3\linewidth]{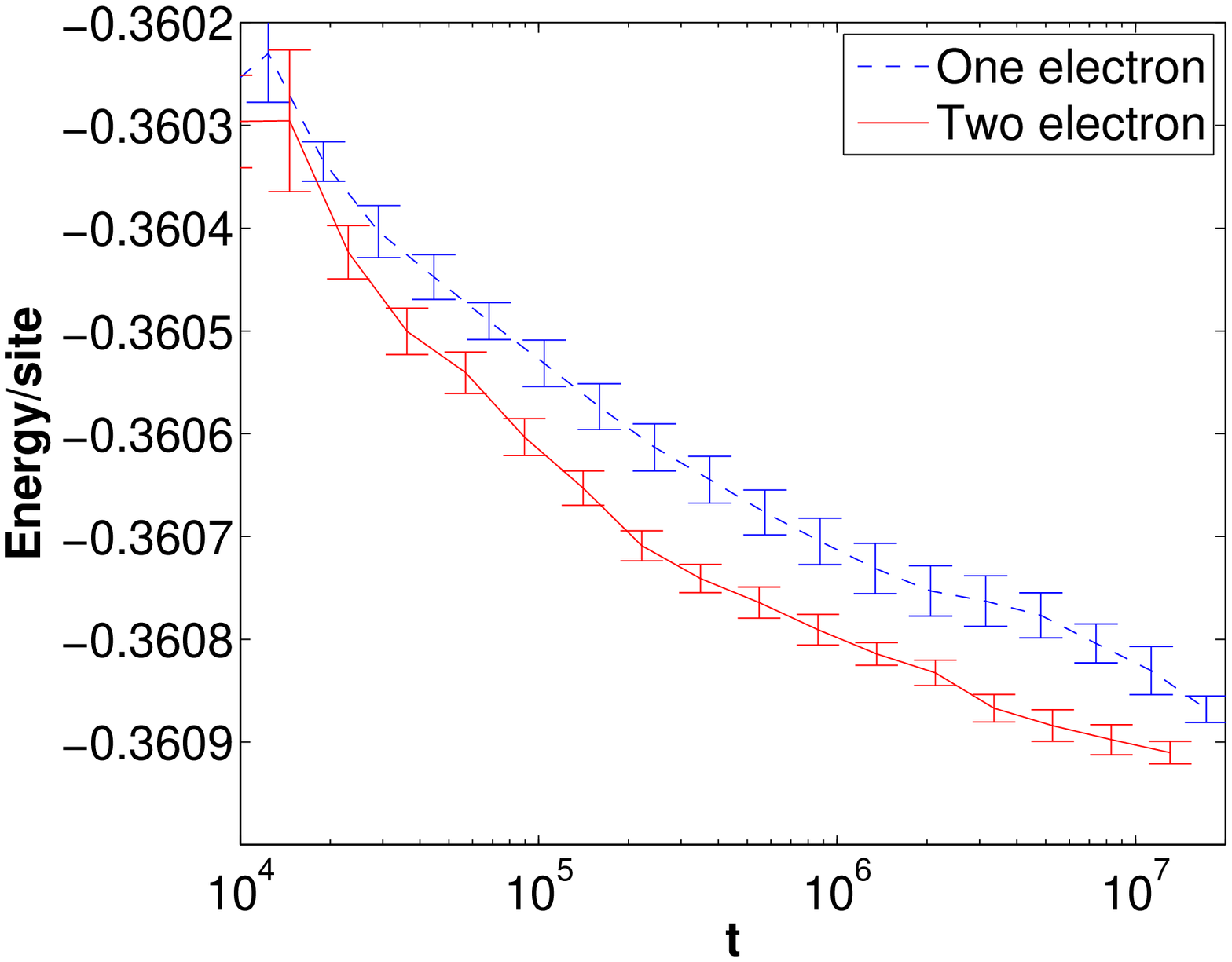}b)    
\includegraphics[width=.3\linewidth]{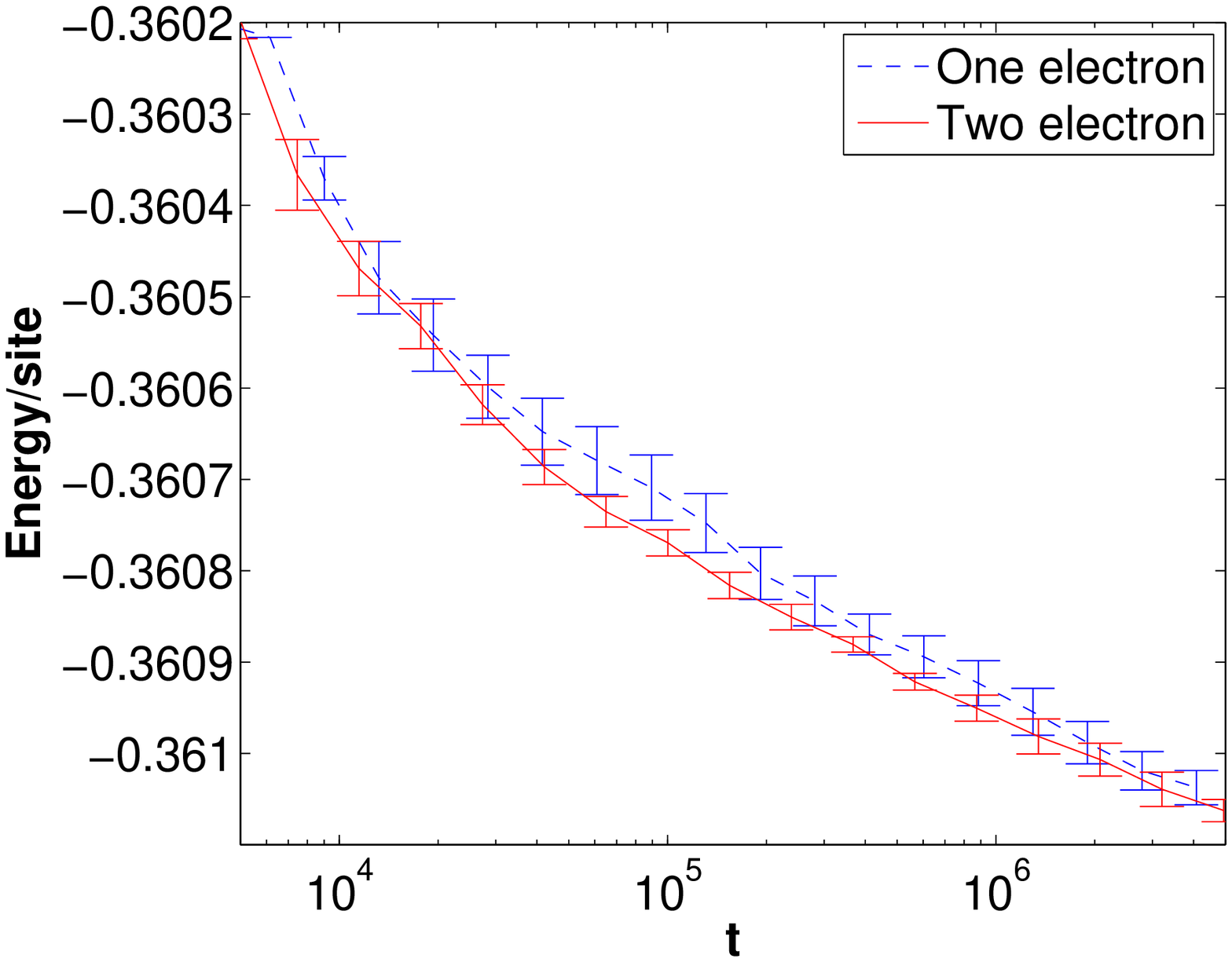}c)
\caption{(Color online) Energy relaxation as function of time at a)
  $T=0.001$, b) $T=0.005$ and c) $T=0.01$. Averages are shown for one and two
  electron transitions, error bars are standard deviation of the mean.
\label{fig:relax}}
\end{figure*}
The results are shown in Fig. \ref{fig:relax}. It is clear that as the
temperature decreases, the difference between the relaxation rates in
the one- and two-electron cases increases.  To confirm that the
results are general and the sample sufficiently large we did one
set of 10 time evolutions on the same sample but starting from a
different initial state  and one set on a different sample.
The same behaviour was seen in all cases.

\begin{figure}
\begin{center}
\includegraphics[width=.95\linewidth]{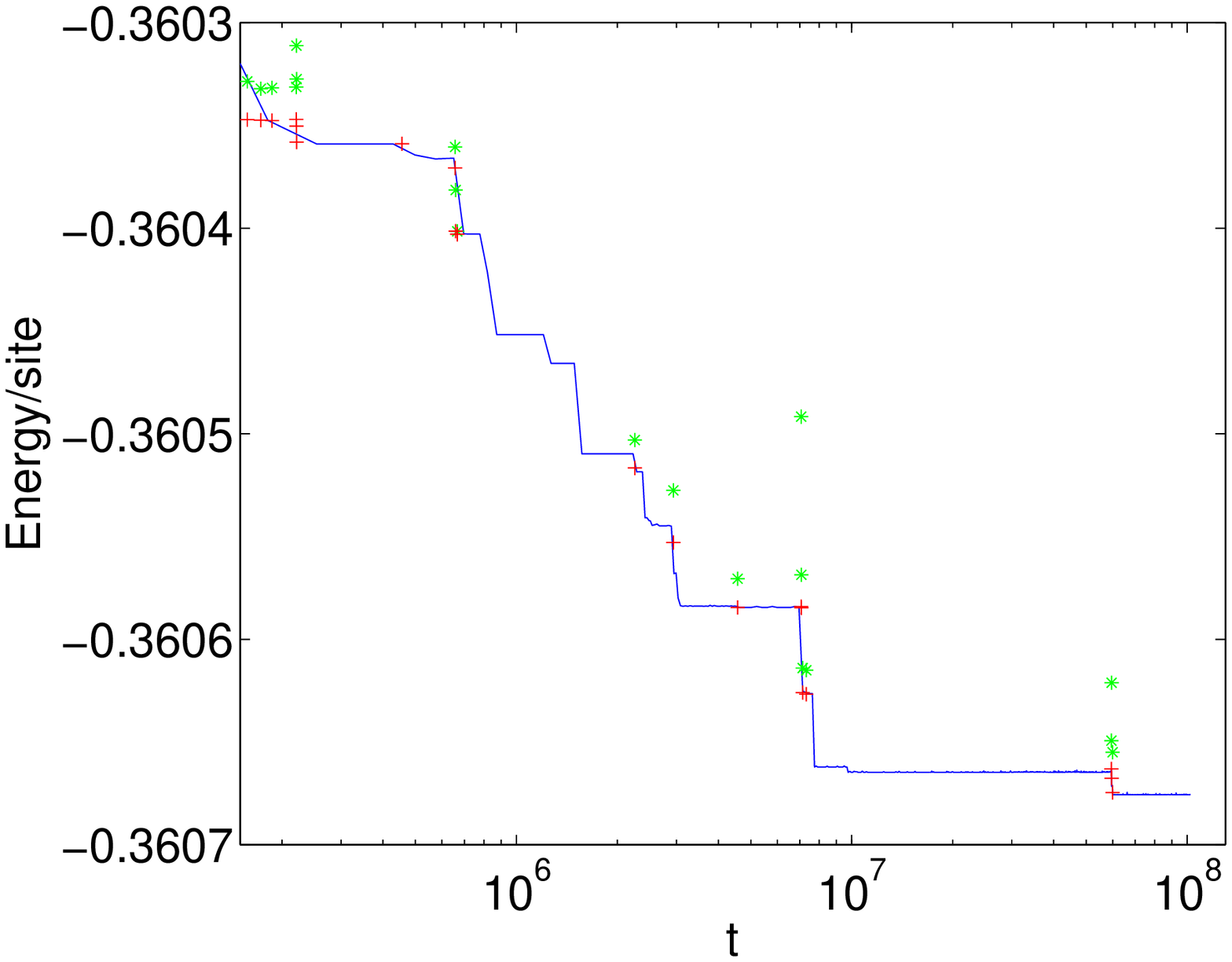}a)   
\includegraphics[width=.95\linewidth]{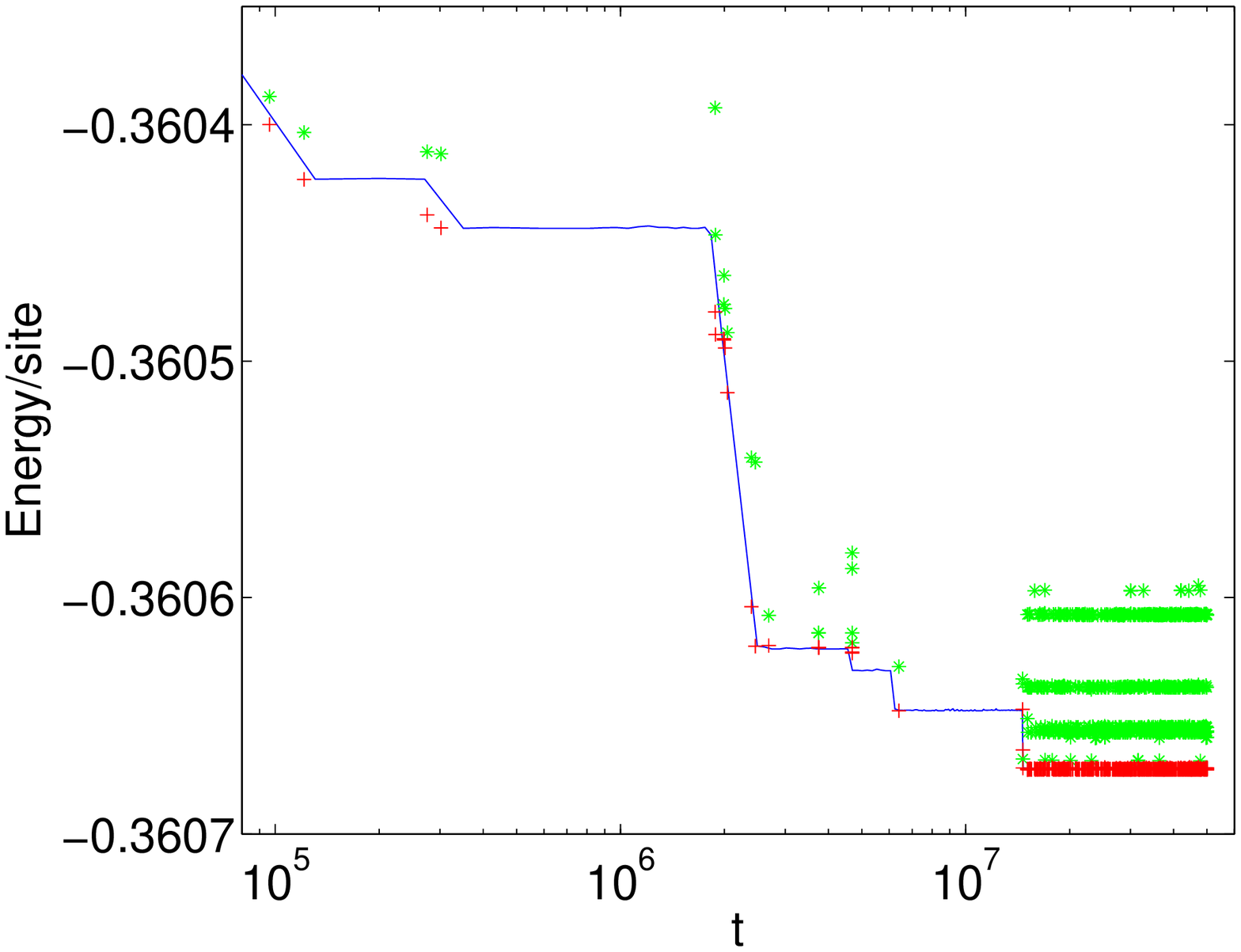}b)   
\caption{(Color online) The correlation between the points where two
  electron jumps occur and steps in the energy relaxation graph. The
  blue curve is the total energy of the system. The red points (+) are the
  final energies after two electron jumps and the green points (*) the
  energy of the intermediate state if this jump was to be replaced by
  sequential one electron jumps. $T=0.001$. The two graphs, a) and b),
correspond to two different Monte Carlo evolution of the same sample
and initial state
\label{fig:correlation}}
\end{center}
\end{figure}

To see more clearly the importance of the two electron jumps we can
look at one particular relaxation graph and mark the points where two
electron jumps occur (Figure \ref{fig:correlation}, the two graphs,
Figure \ref{fig:correlation}a) and Figure \ref{fig:correlation}b)
correspond to two different Monte Carlo evolution of the same sample
and initial state).
In the figure the curve is the energy, while the points mark the time
when a two electron jump was performed. The red points represent the
final energy after the transition, while the green points represent
the energy of the intermediate state if this jump was to be replaced
by sequential one electron jumps. The temperature was
$T=0.001$, and as we can see the increase in energy to the
intermediate state is sometimes more than two orders of magnitude
larger that this. This means that the probability of this process
occurring sequentially is extremely small. As can be seen from the
figure, there are clear correlations between the occurrence of two
electron jumps and steps in the relaxation graph. This means that the
two electron jumps are essential in overcoming barriers in the
relaxation path and give a contribution to the relaxation rate even if
the number of two electron jumps can be a small fraction of the total
number of jumps. Sometimes (at long times in Fig
\ref{fig:correlation}b) there can occur soft two electron excitations
which are then jumping back and forth between the two configurations
like soft dipoles in the one electron case. These give large
contributions if we try to count the number of two electron
transitions, but are not important for relaxation.

\begin{figure}                                                     
\begin{center}
\includegraphics[width=\linewidth]{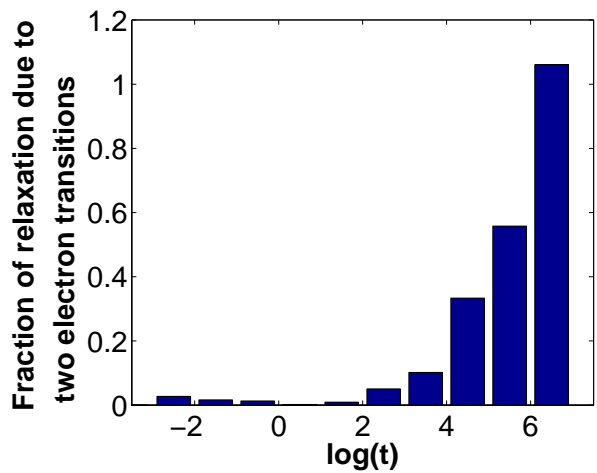}   
\caption{The fraction of relaxation due to two electron
  transitions. The figure is an average over 6 different Monte Carlo 
evolutions on one sample at $T=0.001$. 
\label{fig:fractionOfRel}}
\end{center}
\end{figure}

To better measure the importance of the two electron jumps on the
relaxation, we do the following. For each decade in time we see how
much the energy was reduced in the two electron jumps (or by one
electron jumps immediately following a two electron jump) and compare
this to the total relaxation of energy during this time. We then get
Fig. \ref{fig:fractionOfRel} and as we can see, the two electron jumps
increase in their importance for relaxation, and at the later times in
the simulation, they are responsible for most of the
relaxation. At long times the fraction slightly
exceeds 1 which is due to the fact that the energy of the system was
not stored after each jump so the energy reduction in a two electron
jump can be slightly overestimated if it was preceeded by one electron
jumps which increased the energy. 

\subsection{Relaxation using master equation on the random sites model}

\begin{figure}
\begin{center}
\includegraphics[width=\linewidth]{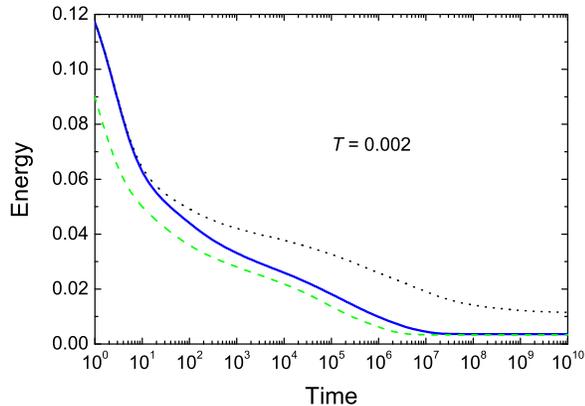}                  
\caption{(Color online) Energy relaxation as function of time for
one-electron jumps (black dotted curve), one- and two-electron jumps
without any prefactor (green dashed curve) and with the prefactor of
Eq.\ (\ref{two}) (blue continuous curve).}
\label{prefactor}                                                         
\end{center}
\end{figure}

We have calculated the average energy, with respect to the ground
state energy, as a function of time. In Figure \ref{prefactor} we
plot the results for one-electron and two-electron relaxations at a
temperature $T=0.002$ for a system with 500 sites. We consider this
small size in order to have configurations extending over a relatively
large energy range.  The dotted line is the result when only
one-electron relaxation is considered.  The continuous and the dashed
curves correspond to relaxation by one- and two-electron jumps. In the
dashed case we have not included any prefactor in the expression for
the transition rates, while in the continuous curve we use Eq.\
(\ref{two}).  We first note how relaxation by one-electron jumps alone
is far from complete.  The system gets easily stuck in metastable
states even for the relatively small system size considered.  The
inclusion of two-electron jumps is almost negligible at short times,
where one can always lower the energy by one-electron jumps. But at
larger times, two-electron transitions are really needed to overcome
the energy barriers.

We also note that if we do not include the right prefactor in the
two-electron rate we are overestimating their effects, specially at
short relaxation times, because we are double counting some
excitations.  We have checked that the results for two-electron
contributions do not change if we only include those transitions with
negative interaction energy. This result in an important reduction in
the number of many-electron excitations to include in the
simulations. In future calculations of many-electron effects it will be
convenient to take advantage of this result and to explore more
drastic reductions in the number of relevant excitations.

\begin{figure}
\begin{center}
\includegraphics[width=\linewidth]{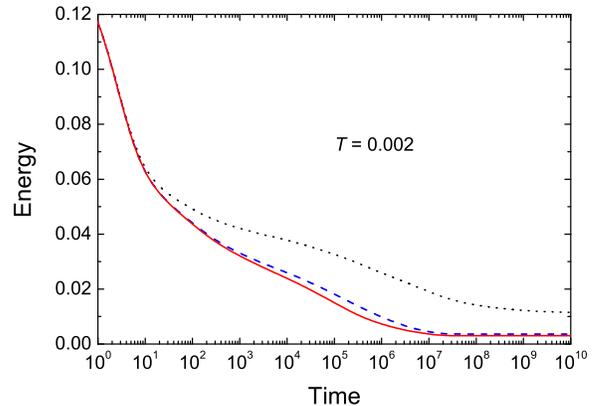}                  
\caption{(Color online) Energy relaxation as function of time for
one-electron jumps (black dotted curve), one- and two-electron jumps
(blue dashed curve) and up to three-electron jumps (red continuous
curve).}
\label{all}                                                         
\end{center}
\end{figure}

At a time of roughly $10^7 \tau_0$ we have already practically
reached the thermal equilibrium when up to two-electron transitions
are considered. We expect this time to increase drastically with
system size.  In figure \ref{all} we have represented energy
relaxation by one-electron jumps (dotted curve), by up to two-electron
hops (dashed curve) and by up to three-electron jumps (continuous
curve).  For three-electron jumps we have used for the prefactor the
sum of all the different products of dipole-dipole contributions. We
note that the inclusion of three-electron jumps does not affect much
the results for the size considered. We expect that their effects will
be more important for larger sizes, which will contain larger energy
barriers and a more complex energy landscape.  As we include
transitions of more particles, the energy relaxation curve seems to
approach a logarithmic behavior.

\section{Results on conductivity}\label{conductivity}

\begin{figure}
\begin{center}
\includegraphics[width=\linewidth]{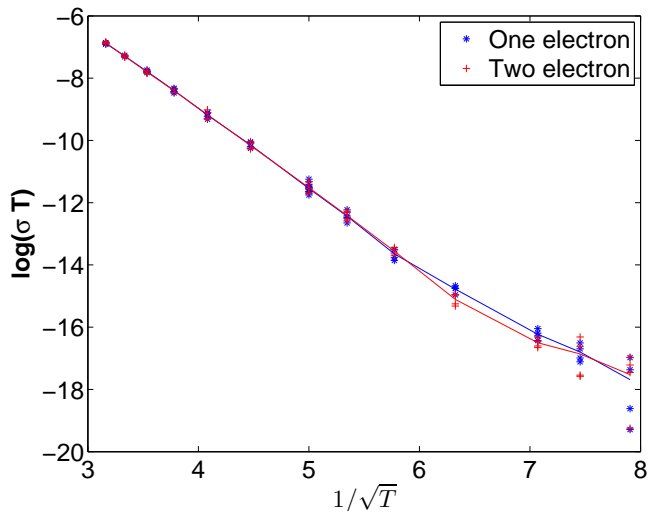}
\caption{(Color online) Conductance as function of temperature showing
  the Efros Shklovskii law. Blue points (*) are with one electron jumps,
  red points (+) include also two electron jumps. Four samples are shown for
  each temperature, the spread indicates the uncertainty. The lines
  are the averages.
\label{fig:ES}}
\end{center}
\end{figure}

We also studied conductivity, comparing the cases with and without two
electron jumps. At temperatures $T\geq$ 0.04
 we ran four different samples of size 100$\times$100. For
$T\leq$ 0.04 we used four
samples of size 200$\times$200 (this because we know that at low
temperatures we see finite size effects in the conductance up to
$L=100$). The electric field was $T/10$, which should be in the ohmic
regime. In each case we ran the simulation for $10^7$ accepted jumps
and checked that this was sufficient to obtain a straight line of
transferred charge as function of time. The conductance is then
given by the slope of this line.  The results are shown in
Fig. \ref{fig:ES}.
We see that we reproduce the Efros-Shklovskii law for the conductivity
and that there is no significant difference when including
two-electron transitions. 

\section{Discussion}

Comparing our results with the previous works
\cite{pg,tsigankov,somoza} it seems that all fall into the same
coherent picture. By focusing on relaxation, we were able to apply the
Monte Carlo method to temperatures comparable to the ones were the
master equation approach can be used. Then we see similar behavior in
both the cases. The energy relaxes faster when two electron
transitions are included. A more direct comparison of the two methods
is difficult since the models differ in several respects.  In the
Monte Carlo method we prefer to use a lattice since it reduces the
computational effort while in the master equation we are resticted to
small samples and prefer a random site model since we believe that
lattice effects are more severe for small systems. Also, the initial
states are different in the two cases, since in the master equation
approach we need to take as initial state some combination of the
states in the low-energy set we are working on. These are already very
low-energy states, and different in structure from the random states
used in the Monte Carlo method. However, we find that our results
convincingly show that both methods give similar results at low
temperatures, and that there are no systematic errors which affect one
or the other method. Furthermore, if we look at Figure \ref{prefactor}
and compare the two graphs including two electron transitions, with
and without the prefactor in Eq. (\ref{two}), we find that although
the omission of the prefactor overestimates the importance of two
electron jumps the results remain qualitatively the same. Thus, the
concern of Tsigankov and Efros\cite{tsigankov} that this
overestimation changes the results qualitatively seems unfounded. We
may then still believe the results of Ref \onlinecite{somoza}, at
least on the qualitative level. Comparing our Figure \ref{fig:ES} with
Figure 2 of Ref. \onlinecite{tsigankov}, both calculations of the
conductivity using the same Monte Carlo method, we find that the two
agree closely (a detailed comparison shows a small shift in the values
but the slope of the line remains the same). Figure 4 of
Ref. \onlinecite{somoza} gives the corresponding results using the
configuration space approach. We see that although the configuration
space method finds a difference in the conductivity when including two
electron jumps, this difference is small, at the level of the
statistical error, for $1/\sqrt{T}\lesssim8$ which is where we have
results using the Monte Carlo method. If we compare with our Figure
\ref{fig:relax}, we see that at these temperatures we can not see any
significant effect of two electron jumps on relaxation either. We
therefore conclude that two electron jumps will only be important at
lower temperatures, and we believe that we would also see this in
Monte Carlo simulations if these could be performed at sufficiently
low temperature.

\acknowledgments

We thank Michael Pollak, Yuri Galperin and Zvi Ovadyahu for useful
discussions.  We also acknowledge financial support from projects
FIS2009-13483 (MICINN), 08832/PI/08 (Fundacion Seneca) and the
Norwegian Research Council. The numerical computations were performed
using the Titan cluster provided by the Research Computing Services
group at the University of Oslo.


\begin{thebibliography}{34}

\bibitem{mott}
N. F. Mott, J. Non-Cryst. Solids {\bf 1}, 1 (1968).



\bibitem{ES}
B. I. Shklovskii and A. L. Efros, Electronic properties of doped
semiconductors (Springer, Berlin, 1984). Section 4.2. 

\bibitem{tenelsen}
K. Tenelsen and M. Schreiber, Phys. Rev. B {\bf 52}, 13287 (1995). 
A. D\'iaz-S\'anchez et. al., Phys. Rev. B {\bf 59}, 910 (1999).

\bibitem{pg}
A. P\'erez-Garrido et al., Phys. Rev. B {\bf 55}, R8630 (1997).


\bibitem{tsigankov} 
D. N. Tsigankov and A. L. Efros, Phys. Rev. Lett. {\bf 88}, 176602 (2002).

\bibitem{somoza}
A. M. Somoza, M. Ortu\~no and M. Pollak, Phys. Rev. B {\bf 73}, 045123 (2006).

\bibitem{zvi}
 A. Vaknin, Z. Ovadyahu, and M. Pollak, 
 Phys. Rev. Lett. {\bf 84}, 3402 (2000).

\bibitem{grenet}
T. Grenet , J. Delahaye, M. Sabra, and F. Gay,
Eur. Phys. J. B {\bf 56}, 183 (2007).


 
\bibitem{BeSo09}
J. Bergli, A. M. Somoza, and M. Ortu\~no, 
Ann. Phys. (Berlin) {\bf 18}, 877 (2009). 

\bibitem{Po81}
M. Pollak, 
Journal of Physics C: Solid State Physics, {\bf 14},
2977 (1981).

\bibitem{ds}
A. D\'{\i}az-S\'anchez, A. M\"obius, M. Ortu\~no, A.
Neklioudov
and M. Schreiber, Phys. Rev. B, {\bf 62}, 8030 (2000).

\end{thebibliography}
\end{document}